\documentstyle[12pt,psfig]{article}
\begin{document}
\setcounter{page}{1}
  \hspace{16.6em}
 {\Large   SUNY -  NTG    95-47
  }
\vspace*{0.8cm}

  \begin{center}
{\bf  \large
Temperatures and Non-ideal Expansion  in
     \\
Ultrarelativistic Nucleus-Nucleus Collisions
       }
\vspace*{0.4cm}\ \\
           H.\ Sorge
 \footnote{
  E-mail: sorge@nuclear.physics.sunysb.edu
          }
\vspace*{0.4cm}\ \\
 Physics  Department,
\vspace*{0.4cm}\ \\
  State University of New York at Stony Brook, NY 11794-3390

  \end{center}

\begin{abstract}
The hadronic phase space distributions calculated with the transport model
RQMD for central S(200 AGeV) on S and Pb(160AGeV) on Pb collisions are analyzed
to study the deviations from  ideal hydrodynamical evolution. After the
preequilibrium stage, which lasts for approximately 4 (2) fm/c in Pb+Pb (S+S)
the source stays in approximate  kinetic equilibrium for  about 2 fm/c at a
temperature  close to 140 MeV. The  interactions  of mesons last until around
14 (5) fm/c during which time strong transverse flow is generated. The
interactions in the hadronic resonance gas are not sufficiently strong to
maintain ideal fluid expansion. While  pions acquire average transverse fluid
velocities around 0.47-0.58 c, heavier particles like protons and kaons cannot
keep up with the pionic fluid,  since their average  velocities are smaller by
about 20 to 30 \%. Although kinetic equilibrium breaks down in the final dilute
stage of $AA$ collisions, the system resembles a thermal system at a
temperature
of 130 MeV, if the free streaming of hadrons after freeze-out is suppressed.
This freeze-out temperature  is consistent with estimates  based on mean free
paths and  expansion rates in a thermal fireball but lower than values
derived from fits to  measured particle ratios and transverse momentum spectra.
The processes in RQMD to which the differences  can be attributed to are the
non-ideal expansion of the hadronic matter and the absence of chemical
equilibrium at freeze-out.
\end{abstract}

\newpage
The dynamical evolution
in ultrarelativistic nucleus-nucleus collisions
proceeds
in three stages.
The initial phase is characterized by mutual interpenetration
 of the two nuclei
which destroys the  coherence of
 ingoing nuclear wave functions and leads to production of
 secondary quanta.
 In the second stage, entropy generation  slows down,
 because the system evolves hydrodynamically, i.e.\
 in or near to local
 kinetic equilibrium.
  The system may be  characterized as a
   hadronic gas or as a quark-gluon plasma
 which  later
  undergoes a phase transition back into the hadronic world.
  Which state is realized,
  depends on the initial conditions, most importantly
  the  energy and baryon density.
  Hydrodynamic expansion cools and dilutes the system
   up to densities at which the hadrons cease to interact (freeze-out).
  Afterwards they
   stream out freely and reach the detectors.
   The extraction of the phase diagram and other
   bulk properties using penetrating probes like
   photons or hadronic final-state observables
   becomes much easier if the  system evolves dominantly according to
   hydrodynamical equations of motions.
  There is probably no doubt that the conditions of hydrodynamical
  evolution can be met during a part of the evolution,  if the
    energies of the colliding nuclei are sufficiently high.
   However, it is worthwile to study whether  hydrodynamical
   behaviour is exhibited in $AA$ collisions at presently
  available energies (up to 200 AGeV at CERN-SPS).
   The hydrodynamical model \cite{CLA86} has been widely used
   to describe nucleus-nucleus collisions in this energy domain
  \cite{GER86}-\cite{SCH93}.
   In particular, it was suggested that the final
    hadron momentum spectra could  be understood as a
   convolution of hydrodynamic motion (collective flow) and a stochastic
    component determined by the freeze-out temperature \cite{LEE89}.
   Subsequently, available experimental data from
   $AA$ collisions at AGS (energy 10-15AGeV) and at CERN
  have been interpreted
   based on these assumptions (see e.g.\
   \cite{VEN90},\cite{SCH92},\cite{BMS95}). However,
  for a satisfactory understanding of the
  experimental results
  it is important
  to  assess how large the corrections to the
   hydrodynamical picture are.

   How non-ideal is the hydrodynamics of  expanding hadronic
    matter which is created in nuclear collisions at
   CERN energy?
    The role of transport coefficients in a thermal hadron gas
   which describe the `restoring forces' in case of
    infinitesimal deviations from equilibrium
   has already been discussed  \cite{LEV91},\cite{PRA93}.
    Here, I  use a semi-classical transport theoretical approach
     (Relativistic Quantum Molecular Dynamics)
   to study the deviations from hadronic fluid dynamics.
   The advantage of a transport calculation is that
    it is not restricted to  situations close to equilibrium.
   In addition to assuming a common flow velocity,
    `model-independent' analysis of
   collective and stochastic component
   by an analysis of experimental
   momentum spectra or correlation functions
   have relied
    on the concept of a thermal state at freeze-out.
  However,
    the assumption of an
   abrupt transition from a system which maintains local
   equilibrium into a gas of free hadrons is a drastic
    idealization.
   In a hydrodynamic description, corrections associated
    with heat conduction and viscosity would start to play
   a significant role before freeze-out.
  Is an analysis of the hydrodynamic flow effect  on the hadron
   momentum spectra jeopardized by the break-down of
    hydrodynamics in the dilute stage of matter evolution?
  In order to address this question
   hadron distributions
  in seven-dimensional phase space
   calculated using the
    RQMD model
    were carefully studied.
  The results of the analysis to be presented will be
  restricted to two systems,
   central Pb(160AGeV) on Pb and
   S(200AGeV) on S collisions. Comparisons between the two will allow us to
   estimate the influence of  finite mass number and size
   on the hydrodynamical behaviour.
    Here, the focus will be on an analysis of the
    transverse degrees of freedom, because
    the expansion in the longitudinal direction is dominated
   by transparency effects as in the  Bjorken scenario \cite{BJO83}.

  The processes which occur in the initial stages
   before hadronization are outside the scope of
  this  Letter.
 In the present context, these processes merely
  serve to set appropriate initial conditions for the evolution
   of the hadronic matter.
  Therefore, I give only a short resume
  of the physics at this stage which is incorporated
  in the transport approach RQMD
   (see Ref.\ \cite{SOR95} for an extensive description).
 The soft -- nonperturbative -- regime of strong interactions
   has been modeled in RQMD as the excitation
  and decay of longitudinally stretched color strings,
   which can be viewed as an
 idealization  of
  chromoelectric flux tubes.
 Flux tubes generated
  in central high energetic nucleus-nucleus collisions
   can be in   higher dimensional charge representations
  of color $SU_3$, because the effective color
   charge per unit transverse area may exceed unity,
   the so-called color rope formation or string fusion.
  I note that the transverse momentum generated
  from rope fragmentation is not much larger than in
   independent string fragmentation, its difference being
   clearly  smaller than the additional transverse momentum
  which is generated after hadronization in the evolution of the
   hadronic resonance matter.
  Furthermore, the initially produced transverse momenta of particles
   in the central
  region
   are oriented   randomly. The collective flow in transverse
  directions starts at velocity
 zero, in contrast to  flow in the beam direction.
  The  hadronization time is aproximately  given by 1 fm/c in S+S
   and 2 fm/c in Pb+Pb  (the latter being larger due to the
    nonnegligible finite size effects related to the
   Lorentz contraction of the ingoing nuclei).

  After hadronization, the hadrons formed are propagated on classical
  trajectories and may interact with each other via binary collisions.
 Mean field  effects, although  built into the RQMD
   model as an option for baryon propagation, are neglected throughout this
  paper.
  Most of their effects in a dilute meson fluid tend to  cancel, because
  their contribution has a  different sign
  above and below  the energy of a resonance pole as noted
 in Ref.\ \cite{SHU90}.
 The most striking feature of hadron-hadron interactions
 at low and intermediate energies, which are relevant after
  hadronization, is the formation of resonances.
 These processes are described by adding
 Breit-Wigner type
  transition rates  (see Ref.\ \cite{SOR95} for details).
 Great care has been exercised to respect detailed balance constraints
  arising from time-reversal invariance,
 in particular with
  resonances  ingoing into collisions.
 Only  then  equilibration studies like the one presented
 here become meaningful.

\begin{figure}[h]
\vspace{-2.0cm}

\centerline{\hbox{
\psfig{figure=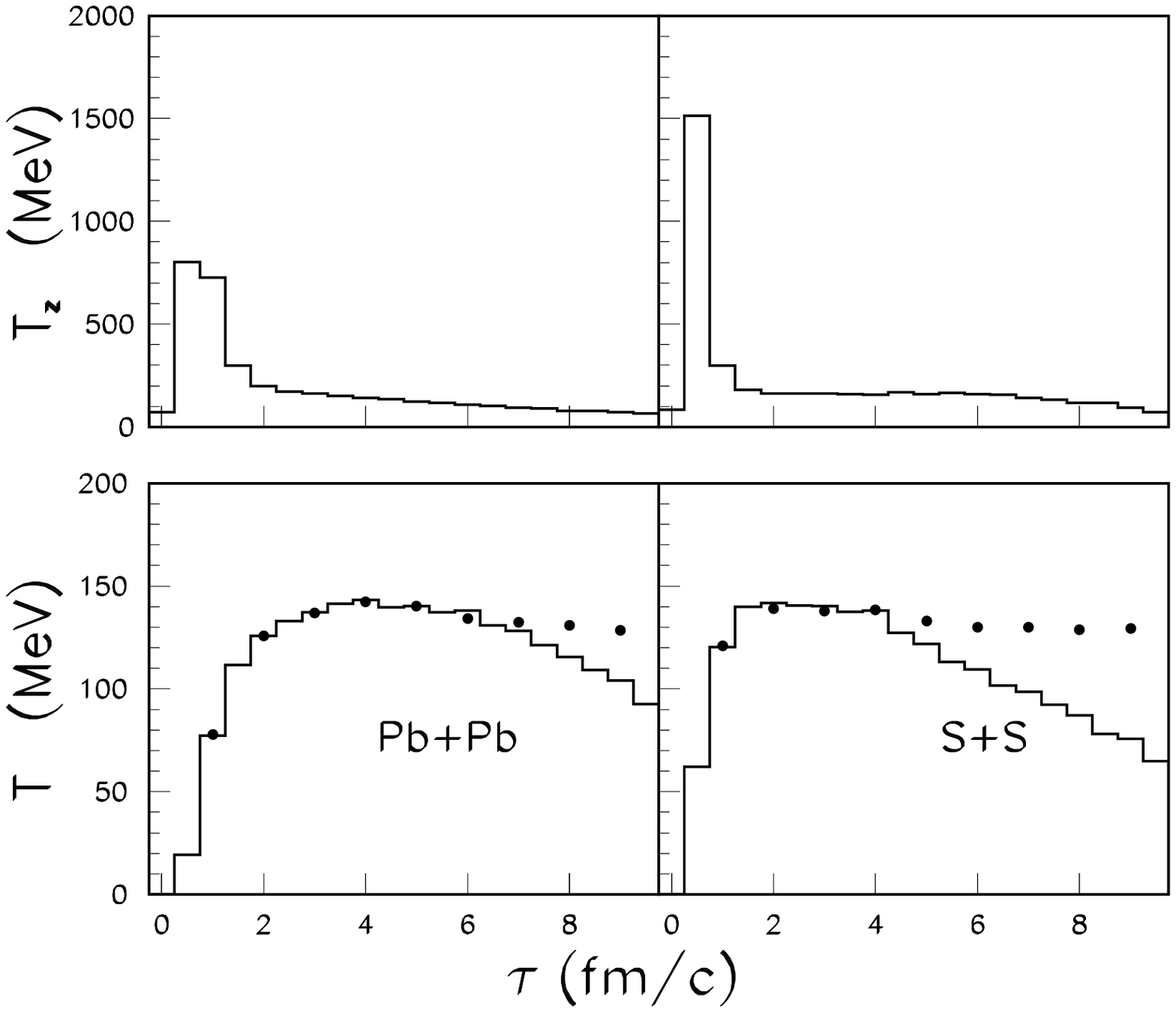,width=20cm,height=20cm}}}

\vspace{-1.0cm}
\caption
[
 ]
{
 \label{aatempev}
}
\end{figure}

 Let us turn now to an analysis of  the space-time evolution
  of the hadronic source as  generated by the RQMD model.
  Fig.~\ref{aatempev} displays the calculated average `temperature' $T$
   in central Pb(160AGeV) on Pb and
   S(200AGeV) on S collisions as a function of
   the boost invariant parameter $\tau$=$\sqrt{t^2-z^2}$,
   with $z$ the coordinate along the beam axis
   and ($t$,$z$)=(0,0) defined as the touching point of the two impinging
   nuclei.
 $T$ is defined here locally by  the ratio between
   the average of the two transverse diagonal components
   of the hadronic energy-momentum  tensor
   and
   the  density of all hadrons, both quantities taken in the
  rest system of the fluid.
  The stress tensor has the standard form given from kinetic theory
 \begin{equation}
   \label{tmunu}
   \Theta^{\mu \nu} =
       \sum _{h}
       \int   \frac{d^3p }{p^0}
         p^\mu   p^\nu
         f_h(x,p)
     \quad ,
 \end{equation}
   with
     $f_h(x,p)$ denoting the one-particle phase space distribution.
  The sum in eq.~(\ref{tmunu})  runs  over all hadron species $h$.
  The spatial average of $T$
  in a hypersurface
  of constant volume
  (chosen as $\pi$~$R_A^2$$\cdot$1 fm)
   with largest local energy density
   is calculated  at fixed $\tau$.
 Of course, in  equilibrium, $T$  just gives
  the temperature of the system.
 The initial strong increase of $T$ with $\tau$ up to 4 (2) fm/c
  in Pb+Pb (S+S) reflects  the preequilibrium
  stage
which is a relic of the flux tube dynamics.
 It  does not imply that a relevant amount of transverse momentum
 per hadron   is created in this stage.
   The  reason for the initially depleted transverse $T$ values is
   that the
   longitudinal hadron momenta which enter
   into the energy denominator of
   eq.~(\ref{tmunu}) start out with an extremely hard distribution.
  The values of  $T_z$  defined by employing the
    longitudinal component of the stress tensor $\Theta ^{zz}$
  show a spike as a function  of $\tau$
  with maxima on the order of  1 GeV
   (cf.\ fig.~\ref{aatempev}).
 Ingoing nucleons which have not
  collided yet are excluded
   in calculating the stress tensor and  hadron densities.
  In terms of an effective equation of state the transverse
  pressure $P$ is `ultra-soft' as a function of energy density
   $\epsilon$, the effective  `bag constant' defined
   by   ($\epsilon$--3$P$)/4 well in excess of (200 MeV$)^4$
   at this stage.
 The maxima of the function $T$($\tau$)
  close to 4 (2) fm/c are  reached at a time  when
  the longitudinal $T$ values approach the corresponding transverse
 values, and the local momentum distributions become isotropic.
 Only after this has happened can local equilibrium  be
 achieved, and a hydrodynamic concept will  give
  equivalent results. Both in Pb+Pb and in S+S,
 the system stays at a  temperature near 140-130 MeV for about 2 fm/c.
 The temperature range
  found from the calculations can be considered as an
 a posteriori justification of the
  approach taken in RQMD which combines 1+1 dimensional
   prehadronic (flux tube) dynamics with the resonance gas
  picture for the hadronic stage.
 It is expected that
 a system with
 temperatures of this order can be reasonably well described in
 terms of hadronic degrees of freedom \cite{BEB92}.

\begin{figure}[h]
\vspace{2.0cm}

\centerline{\hbox{
\psfig{figure=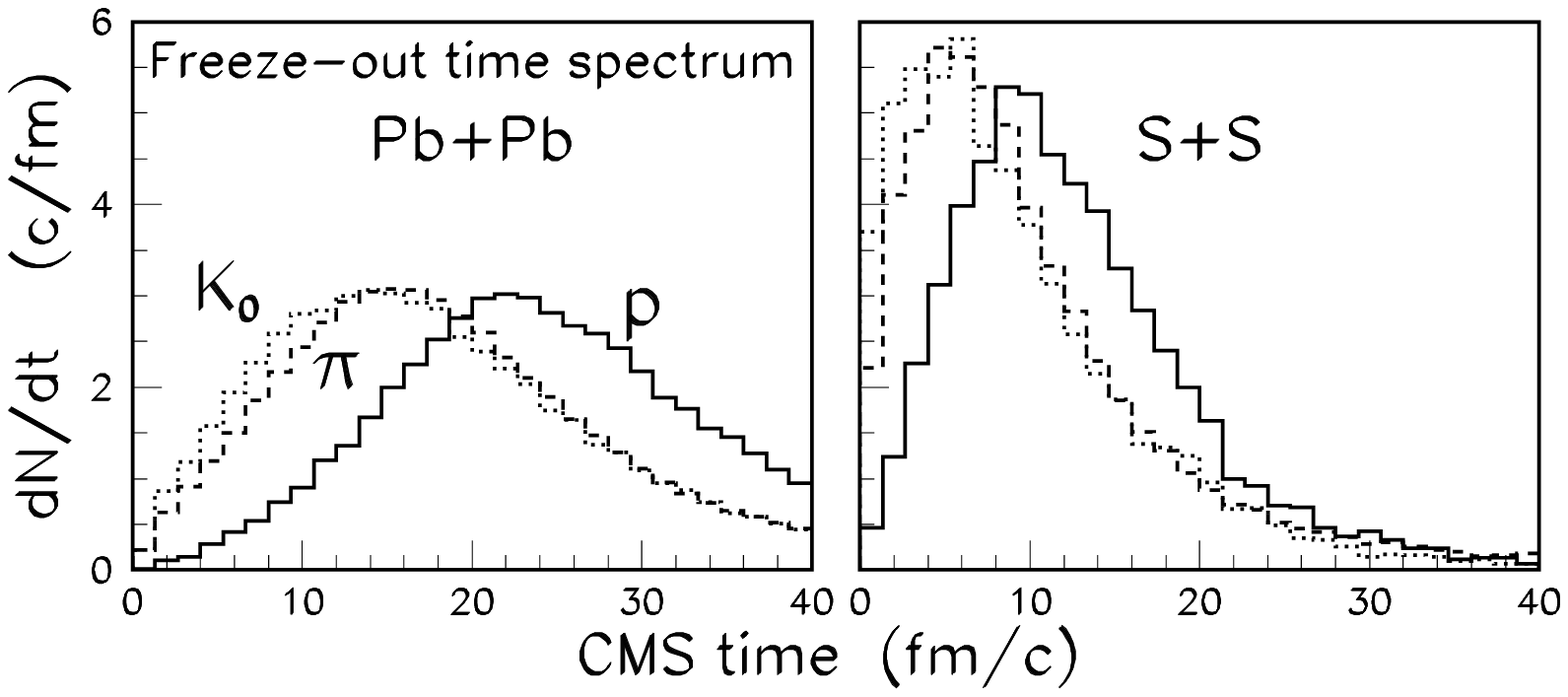,width=20cm,height=20cm}}}

\vspace{-7.0cm}
\caption
[
 ]
{
 \label{aadndt}
}
\end{figure}

 After some time, $T$  starts to
  drop continuously in Pb+Pb (S+S) reactions,
  far below values usually assumed
  for the freeze-out temperature  in hydrodynamical
  calculations which are based on
   mean free path arguments
  \cite{BEB92}. Indeed, the strong drop seen in the
  RQMD calculation is related to the onset of hadronic freeze-out.
  In   fig.~\ref{aadndt}
  the calculated spectrum of times in the C.M.\ frame
  at which hadron of different species
  (protons, pions and kaons) have their last interaction
  (collision, decay)
  is displayed.
  The maxima of these freeze-out time distributions
  are  around 14 (5) fm/c for mesons,
  but the distributions are rather wide.
  A  considerable width
  of the freeze-out hypersurface in the time-like direction
  is also found in other transport calculations
   (see e.g.\ Ref.\ \cite{BRA95} for a discussion  of
   $AA$ collisions at AGS energies) and may eventually be
  included in hydrodynamic calculations  \cite{GRA95}.
  Continuous decoupling of hadrons
    influences the cooling
   process  in the high energy density region
  as early as after 6 (4) fm/c.
  The calculation of the  hadronic stress tensor
   has been repeated,  artificially
     suppressing the free streaming after freeze-out by
     `glueing' the hadrons to the position
   of their last interaction.
   This modified analysis can be employed
    to  estimate
   the additional
   cooling effect due to the
   freely streaming component in the gas.
  The result for the modified  $T$ evolution is also presented
   in fig.~\ref{aatempev} (shown by dots). It demonstrates
   that the drop of the $T$  values with time
    is  caused
    by  free hadron streaming after freeze-out.

  By construction, $T$ calculated in the modified analysis
   has to  approach a constant value
  after  all strong interactions have ceased.
   One can see that $T$ approaches 130 MeV in both reactions,
   if  free streaming  is suppressed.
   It is reasonable to compare this asymptotic $T$ value
    to the  freeze-out temperature in hydrodynamical
   calculations with instantaneous conversion of
     local fluid elements into free hadrons \cite{COO71}.
  A freeze-out value of 130 MeV for a fireball of a few fermi size
   is  consistent with
   the criterion that
   the collision rate
   should be  smaller
   than the expansion rate
    in order that hydrodynamics be applicable.
   Applying this criterion to a baryon-free fireball the authors of Ref.\
  \cite{BEB92} find a freeze-out temperature
  around 140 MeV and correspondingly smaller values if a positive
   chemical potential for pions is built up.

 Some remarks are in order  with respect to recent interpretations
  of experimental data for central S on A collisions at  200 AGeV.
  The temperatures which have been fitted to final
  particle ratios  either in a hadronic gas \cite{DAV91,CLE93} or in a
  deconfined plasma breakup scenario \cite{RAF91} (however, without
  the contribution from gluon fragmentation)
  are  considerably larger (in the range 160 to 230 MeV)
  than the  freeze-out `temperature' value found here.
  This may not be really surprising, because chemical equilibrium
 is expected to be lost much earlier than kinetic
  equilibrium \cite{BEB92}. The RQMD calculations support this
  scenario.
  In particular, the strong
   strangeness enhancement in the (anti-) baryon sector
  experimentally observed
   by the  NA35 and WA85 groups \cite{QM93}
   which calls for temperatures in excess of 160 MeV
   in chemical equilibrium approaches
   is well reproduced by RQMD \cite{SOR92}.
   From the RQMD calculations, I therefore conclude  that
  chemical and kinetic equilibrium
  are not simultaneously attained  in a sizable fraction of the  source.

\begin{figure}[h]
\vspace{-3.0cm}

\centerline{\hbox{
\psfig{figure=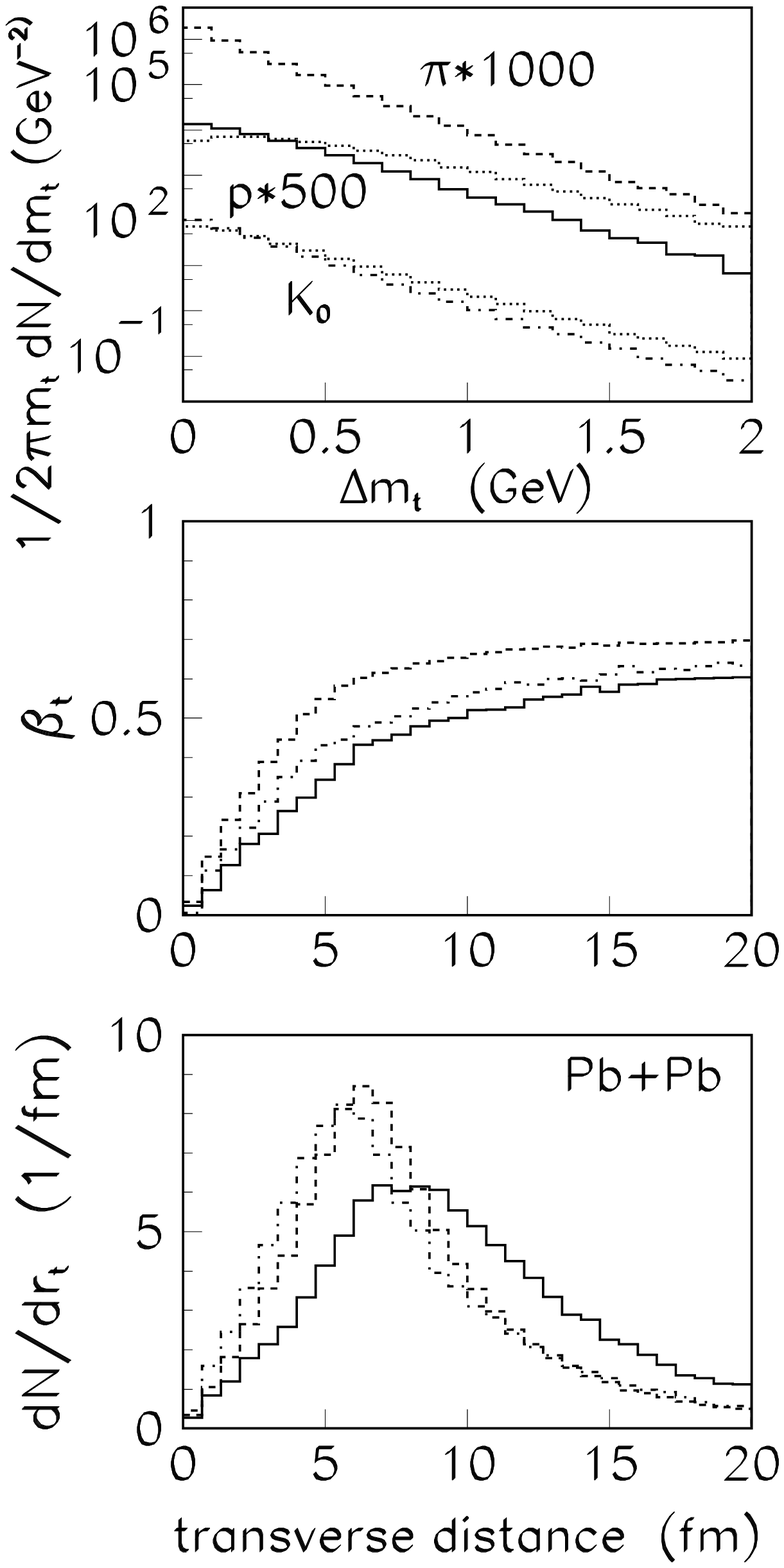,width=10cm,height=20cm}}}

\caption
[
 ]
{
 \label{pbpbtrdflm}
}
\end{figure}

  The final hadron transverse mass spectra  measured
   for S induced reactions
   exhibit an approximately exponential behaviour with
   rather similar inverse slope parameters (`apparent temperature')
   around 200  MeV in central S+S reactions
   and 230 MeV for heavy targets. Pion spectra
   for which concave shapes have been observed are the only exception.
  The Boltzmann-type  spectra  make it
   difficult to disentangle flow and temperature effects unambiguously
   \cite{VEN90,SCH92}. Smaller freeze-out temperatures can be traded against
   larger transverse flow velocities and vice versa.
   It was suggested
   in Ref.\ \cite{SCH92}  that the measured spectra allow
     average flow velocities  of
    at most 0.4 c in S+S collisions.
     `Circumstantial evidence' for a freeze-out scenario
    with $\langle \beta _t \rangle $=0.25 c and
    $T$=150 MeV was presented. However,
   RQMD calculations
   whose results agree  rather well with the experimental data
   yield  even  larger
   collective velocities.
   Averaged over the
   freeze-out hypersurface,
   $\beta_{t}$
   of pions in the central rapidity region ($y_{CMS}$$\pm$1),
    amounts to 0.47 for  S+S and 0.58 c for Pb+Pb.
  On the other hand, particles heavier than pions
   exhibit less flow: e.g.\ for protons
    $\beta _t$=0.41 (0.46) and for kaons $\beta _t$=0.34 (0.45)  in S+S
    and  Pb+Pb collisions respectively.
   The observed effect is in line with
    the findings in Ref.\ \cite{PRA93} that
   more massive hadrons than  pions equilibrate  slower
   at relevant temperatures around 150 MeV, in particular have
   much larger energy relaxation times.
   Therefore these hadrons
    cannot be `dragged' by  a  fluid composed mostly of pions.
  As a result,
   particles with larger rest mass  experience smaller flow
   velocities,
  but their momenta are more sensitive to the boost from flow.
   An analysis of the resulting momentum spectra
   based erreonously on the assumption of a common flow velocity will
   tend to overestimate  the freeze-out temperature.
  This effect is  demonstrated in fig.~\ref{pbpbtrdflm}
   in which the transverse mass spectra of pions, protons and kaons
  at central rapidity are shown together with the average
    transverse velocities and densities at freeze-out as a function of
   transverse distance
  (for Pb(160AGeV) on Pb).
   Fitting the spectra by exponentials, the  pion slope parameter has
   a value of 230 MeV
   at large transverse masses $>$$m_0$+1  GeV/c$^2$, very similar
   to proton and kaon slopes.
  Fig.~\ref{pbpbtrdflm} also contains modified spectra for protons and
  kaons in which  these particles get an additional transverse boost
   with
 $\Delta$$\beta _t$=min($\beta_{max}$,~($r_t$/ 3fm) $\cdot \beta_{max}$)
  to mimic the larger flow velocity of pions
 ($\beta_{max}$=0.2/0.15 c for $p$/$K$).
  It is obvious that an  interpretation of the inverse slope as a
  common temperature would become impossible, if proton
  and kaon distributions had  the same velocity profile as the pion fluid.
   Transverse mass distributions of
    nuclear clusters like deuterons may
    clarify the role of collective flow  in the
   forthcoming Pb on Pb experiments at CERN-SPS.
   RQMD  predicts
   a pronounced shoulder-arm shape  of the spectrum
   which is characteristic of collective flow \cite{MAT95}
    and  is missing  in S induced collisions
   in case of truely heavy ion reactions
   \cite{SOR95B}.

In summary,
the hadronic phase space distributions calculated with the transport model
RQMD for $AA$ collisions at 160-200 AGeV  have been analyzed.
After  a preequilibrium stage, the hadronic matter starts to expand
into the transverse directions, with a temperature of around 140 MeV.
While the pions acquire average transverse fluid velocities
of up to 0.6 c, heavier particles like protons and kaons cannot
 keep up with the pionic fluid, having  average
 velocities lower by about 20 to 30 \%.
 Although kinetic equilibrium breaks down in the final
 dilute stage of $AA$ collisions, the  system
  locally resembles a  thermal system at temperature 130 MeV
 if the free streaming of final-state hadrons
 is suppressed.
  This freeze-out temperature  is consistent with estimates
  based on mean free paths and
   expansion rates in a thermal fireball but lower
  than values suggested in the literature  which were derived
  from fits to
  measured particle ratios and transverse momentum spectra.
 The processes  in RQMD to  which the differences
 can be attributed to are the non-ideal expansion of the hadronic matter
  and the absence of chemical equilibrium at freeze-out.

\newpage

 {\noindent   \bf
  Acknowledgements
 }

  The author would like to thank the
  theory group of the Gesellschaft f.\
   Schwerionenforschung (Darmstadt)
    for kind  hospitality
     during his visit from July until September.
  He thanks M.\ Prakash for a careful reading of the manuscript.

\newpage

{\noindent  \LARGE   Figure Captions:}
\vspace{0.6cm}

{\noindent \large Figure 1: }

{\noindent
Evolution of $T$  as a function of $\tau$:
 $T$ is defined by the ratio of one of the spatial diagonal components
  of the hadronic stress tensor $\Theta ^{ii}$
  and the hadron density, both quantities evaluated in the local
  rest system of the hadronic matter. $T$ is  equal to the temperature
   for a classical ideal gas in equilibrium.
  The upper (lower) part of the figure displays the evolution of $T$
   employing the longitudinal
   (average of the transverse) component(s) of $\Theta ^{ii}$.
   Stress tensor components and hadron density are averaged over
    the region of highest local energy density with total volume
   $\pi$~$R_A^2$$\cdot$1 fm in the CM frame.
  The RQMD (version 2.1) results
  for the system Pb(160AGeV) on Pb are shown
  on the left, for S(200 AGeV) on S on the
   right hand side.
   Both reactions have been calculated for impact parameters
  less than 1 fm.
   The results of the default calculation are represented by
   histograms. Also shown is the result (dots)
  which has been obtained
   by freezing the  3-vector positions of  hadrons
   after they have suffered their last interaction.
}
\vspace{0.2cm}

{\noindent \large Figure 2: }

{\noindent
 The distribution of freeze-out times evaluated in the CM frame
  for different hadron species:
  protons (straight lines), pions (dashed lines) and
   neutral (anti-) kaons (dotted lines).
   Only particles with CM rapidity less than one  are included.
  The RQMD results
  for the system Pb(160AGeV) on Pb are shown
  on the left, for S(200 AGeV) on S on the
   right hand side.
  (The distributions have been renormalized to give the same
  integral as for the protons in Pb+Pb.)
}
\vspace{0.2cm}

{\noindent \large Figure 3: }

{\noindent
  Transverse mass spectra 1/2$\pi$ d$N$/$m_t$d$m_t$
   as a function of $\Delta$$m_t$=$m_t$-$m_0$ (top),
  average transverse velocity  $\beta _t$  (middle)
   and density distribution d$N$/d$r_t$  (bottom)
   which are integrals over the freeze-out hypersurface
   but keeping the transverse freeze-out distance $r_t$ fixed.
    The results are
    for protons, neutral kaons and pions  ($\pi^+ $+$\pi ^-$/2)
  in a central rapidity window
  ($y_{CMS}$$\pm$1).
   The calculations have been done with RQMD
  for the system Pb(160AGeV) on Pb with
   $b<$1 fm.
  (The freeze-out density
   distributions of kaons and pions have been renormalized to give the same
  integral as for the protons.)
}

\newpage

\newpage

\newpage

\end{document}